\documentclass[11pt]{article}
\usepackage{amsmath, amssymb, amscd, amsthm, amsfonts,bm}
\usepackage{graphicx, float}

\usepackage{bibentry}
\usepackage[english]{babel}
\usepackage[natbibapa]{apacite}
\usepackage[colorlinks=true, allcolors=blue]{hyperref} 

\usepackage{authblk}

\author[1]{Zihan Zhou\thanks{These authors contributed equally to this work.}}           
\author[1]{Zizhong Tian$^*$}              
\author[2]{Christine B.\ Peterson}      
\author[3]{Le Bao}                      
\author[1]{Shouhao Zhou\thanks{Corresponding author. Email: szhou1@pennstatehealth.psu.edu}}       

\affil[1]{Department of Public Health Sciences, The Pennsylvania State University College of Medicine, Hershey, PA 17033}
\affil[2]{Department of Biostatistics, The University of Texas MD Anderson Cancer Center, Houston, TX 77030}
\affil[3]{Department of Statistics, The Pennsylvania State University, University Park, PA 16802}

\title{Shiny-MAGEC: A Bayesian R Shiny Application for Meta-analysis of Censored Adverse Events}
\date{}

\begin{document}
\maketitle

\begin{abstract}
Accurate assessment of adverse event (AE) incidence is critical in clinical cancer research for drug safety evaluation and regulatory approval. While meta-analysis serves as an essential tool to comprehensively synthesize the evidence across multiple studies, incomplete AE reporting in clinical trials remains a persistent challenge. In particular, AEs occurring below study-specific reporting thresholds are often omitted from publications, leading to left-censored data. Failure to account for these censored AE counts can result in biased AE incidence estimates. 
We present an R Shiny application that implements a one-stage Bayesian meta-analysis model specifically designed to incorporate censored AE data into the estimation process. This interactive tool provides a user-friendly interface for researchers to conduct AE meta-analyses and estimate the AE incidence probability following the bias-correction methods proposed by \citet{xinyue2024meta}. It also enables direct comparisons between models that either incorporate or ignore censoring, highlighting the biases introduced by conventional approaches. This tutorial demonstrates the Shiny application’s functionality through an illustrative example on meta-analysis of PD-1/PD-L1 inhibitor safety and highlights the importance of this tool in improving AE risk assessment. Ultimately, the new Shiny app facilitates more accurate and transparent drug safety evaluations. The Shiny-MAGEC app is available at: \url{https://zihanzhou98.shinyapps.io/Shiny-MAGEC/}.

\end{abstract}

\section{Introduction}
Meta-analysis plays a crucial role in assessing drug safety (or equivalently, quantifying drug-related harms) \citep{us2018guidance,wang2019treatment}. By synthesizing evidence from multiple studies, it enhances the precision of risk estimates for adverse events (AEs), which is essential for a comprehensive understanding of a drug’s safety profile. Since individual clinical studies are often underpowered for detecting rare harm signals, meta-analysis serves as a valuable tool to combine data across studies, increase sample size, and improve power, enabling a more robust evaluation of drug-related risks \citep{stoto2015drug, hong2021meta}.

However, a major challenge in safety meta-analysis is the incomplete reporting of AEs in cancer clinical trials, which can lead to biased estimates if not properly addressed \citep{huang2011pitfalls,xinyue2024meta}. While the NCI Common Terminology Criteria for Adverse Events (CTCAE) v5.0 catalogs over eight hundred unique AEs \citep{CTCAE5}, trial publications typically report only the most common or severe, while rare or less frequent events may be omitted due to space constraints or study-specific reporting thresholds. 
For instance, when extracting AE count data for grade 3 or higher pneumonitis for PD-1 and PD-L1 inhibitors, researchers may not find explicit AE frequency reporting. Instead, they may see a footnote in the main text or supplementary materials stating that AE counts were only included and reported if they exceeded a certain proportion of the study sample size \citep{powles2018atezolizumab, nanda2016pembrolizumab}. In many oncology trials, AEs may only be listed if they occur in more than $5\%$ of patients, resulting in \textit{left-censored} AE data. Indeed, left-censoring is a common incomplete AE reporting phenomenon, as demonstrated by discrepancies between published articles and unpublished trial documents \citep{golder2016reporting}. Moreover, reporting cutoffs are sometimes defined separately for different severity grades (for example, a non-reporting threshold for all-grade AEs and a separate threshold for grade 3 to 5 AEs) \citep{qi2022bayesian}. This variability in reporting practices highlights the difficulty of harmonizing AE data in meta-analysis. 

Failure to account for left-censored AE data can markedly bias meta-analysis results. One na\"ive approach is to treat an unreported AE as if it did not occur in that study, which effectively imputes a zero count for missing categories. This approach allows the inclusion of all relevant studies in the meta-analysis and is straightforward to implement. However, the assumption can be problematic since the absence of an AE in a report doesn’t always mean it didn’t occur. If an AE was unreported only because it fell below the reporting threshold, assuming it to be zero underestimates the true event rate and can paint an overly safe picture of the drug. 

Another common na\"ive approach is to omit data from studies for an AE outcome if that AE wasn’t reported, equivalent to performing a complete-case analysis for data with missingness. Essentially, the meta-analysis is run only on the subset of studies that provided data for that outcome. While this avoids making false zero assumptions, it introduces its own bias. If unreported AEs are ignored in a meta-analysis and only the studies with higher observed AE counts are included, the estimated incidence rates will be skewed upward, since only studies with higher event occurrences contribute \citep{grigor2019risks,zhou2021treatment, fujiwara2024treatment}. Additionally, complete-case analysis wastes the important partial information in those omitted studies where the event count was rare enough to not be reported, and it reduces the overall sample size and power of the meta-analysis.  

Subsequently, both of these na\"ive strategies distort the true safety profile of treatments. If AEs are under-estimated, patients might be unknowingly exposed to risks; conversely, overestimating AEs could deter the use of an effective treatment, emphasizing the need for appropriate statistical approaches tailored to handle censored AE data. Notably, standard methods for missing data in meta-analyses  \citep{higgins2008imputation,white2008allowingB,mavridis2014addressing} are not designed or suitable to address this specific form of missing-not-at-random data arising from incomplete AE reporting.

To address this pervasive gap in the meta-analysis of safety data, \citet{xinyue2024meta} developed a Bayesian meta-analytic model, named MAGEC (\underline{m}eta-analysis of \underline{a}dverse dru\underline{g} \underline{e}ffects with \underline{c}ensored data), which can correctly handle the censored AE data and provide exact inferences for a balanced, evidence-based understanding of drug harms. Specifically, the censoring cutoff information is incorporated into the likelihood function to enhance the estimation of AE incidence probabilities. Rather than imputing missing counts externally, their work treats unreported AEs as latent variables; the Bayesian model then elegantly integrates over the uncertainty of unreported AEs and draws inference on those hidden counts as part of the Markov Chain Monte Carlo (MCMC) estimation while simultaneously accounting for model between-study variability. As a result, this approach was shown to effectively produce unbiased estimates of AE incidence. 

In this work, we develop an R Shiny application that implements the Bayesian MAGEC model in a user-friendly, interactive platform. This tool provides applied researchers and clinicians with a practical means to conduct meta-analyses of AEs in the presence of incomplete reporting. With the app, users can input aggregated AE data (including the information on reporting cutoffs) and obtain posterior estimates of AE incidence under the Bayesian model. Additionally, the application also allows result comparison between models that account for censoring versus those that ignore it, thereby directly illustrating the bias that can arise from na\"ive complete-case analysis. By packaging advanced methodology into an accessible Shiny interface, our tool lowers the barrier for broader adoption of these statistical best practices in routine safety evidence synthesis.

This paper is organized as follows. Section \ref{sec:model} provides a brief overview of the Bayesian MAGEC model proposed by \citet{xinyue2024meta}. Section \ref{sec:app} presents the functionalities of the R Shiny application. Section \ref{sec:example} offers an illustrative example using clinical trial data on PD-1/PD-L1 inhibitors to demonstrate the app in practice. Finally, Section \ref{sec:discussion} concludes with a discussion of key takeaways, the potential impact of this tool on drug safety evaluation, and future directions for research and practice.

\section{A Brief Review of the Bayesian MAGEC Model} \label{sec:model}
To address the challenge of left-censored adverse event (AE) data in meta-analysis, \citet{xinyue2024meta} developed MAGEC, a Bayesian random-effects meta-analytic model that accounts for censored AE outcomes. It explicitly accounts for censoring by incorporating study-specific reporting thresholds into the likelihood function, enabling more accurate estimation of AE incidence rates while appropriately reflecting the uncertainty introduced by missing data. 

Let $Y_i$ denote the count of AEs under a target severity interval (e.g., all-grade or grade 3 and above) in study $i$, and $ N_i $ represent the number of patients treated in that study. $Y_i$ is assumed to follow a binomial distribution:
\[
Y_i \sim \text{Binomial}(N_i, \theta_i),
\]
where $ \theta_i $ represents the study-specific incidence probability. To account for left-censored AE data, the full likelihood function integrates both completely observed and censored data: 
\[
\mathcal{L} 
= \prod_{o=1}^{O} f_Y(y_o) \prod_{l=1}^{L} F_Y(c_l) 
= \prod_{o=1}^{O} f_Y(y_o) \prod_{l=1}^{L} \sum_{k_l=0}^{c_l} f_Y(k_l), 
\]
where $O$ represents the sets of studies with fully-observed AE counts, and $L$ represents the set of studies where the AE count is left-censored below a study-specific cutoff $c_l$. The censored probability component, $F_Y(c_l)$, ensures that the analysis incorporates information about studies where AEs were unreported due to cutoff-based omission rather than true absence.

The incidence probability parameter $\theta_i$ can be modeled based on the link function $g(\cdot)$ that can transform values in the probability space into the real-value space. In the Shiny application, a logit link was used, such that
\begin{align*}  
g(\theta_i) = \text{logit}(\theta_i) = \mu + \alpha_i + \mathbf{X}_i \boldsymbol{\beta},
\end{align*}
where $\mu$ is the overall log-odds of AE incidence, $\alpha_i$ are the study-specific random effects, $\mathbf{X}_i$ is the design matrix of study-level covariates in the $i$th study, and $\bm{\beta}$ are the effect parameters corresponding to the study-level factors. In the current version of the Shiny application, specifications of study-level covariates are not available, and $\alpha_i$ represents the study-specific random effects. The $\alpha_i$ parameters follow conditionally independent normal distributions with mean 0 and variance $\tau^2$, where $\tau$ quantifies the between-study heterogeneity of the log odds. A noninformative normal prior distribution $\mathcal{N}(0, v_0^2)$ is placed on the overall mean parameter $\mu$ ($v_0^2=10^4$ by default). A half-Cauchy prior distribution $C^+(0,A)$ is placed for the between-study variance parameter $\tau$. 
The half-Cauchy prior with its scale parameter $A$ equal to 2.5, 10, or 25 was recommended in \citet{gelman2008weakly} for the variance parameters in logistic regression models. The estimation for posterior inference is implemented using Just Another Gibbs Sampling (JAGS) \citep{plummer2003jags}, and carried out using a data-augmentation strategy for censored data  \citep{qi2022bayesian}.

This MAGEC model mitigates bias in incidence estimation while incorporating between-study variability. The hierarchical Bayesian framework avoids relying on asymptotic normality for inference, improves estimation stability through MCMC sampling \citep{hamza2008binomial}, and enhances the reliability of drug safety assessments. Simulation studies have demonstrated its robustness in small-sample scenarios and for rare events.

\section{The R Shiny Application: Shiny-MAGEC} \label{sec:app}
We have developed an R Shiny application named Shiny-MAGEC that implements the Bayesian meta-analysis model reviewed in Section \ref{sec:model}. Users can upload their raw AE data collected from multiple clinical studies to obtain meta-analytic estimates of the overall AE incidence probability and the between-study heterogeneity. 

\begin{figure}[h!]
    \centering
    \includegraphics[width=\linewidth]{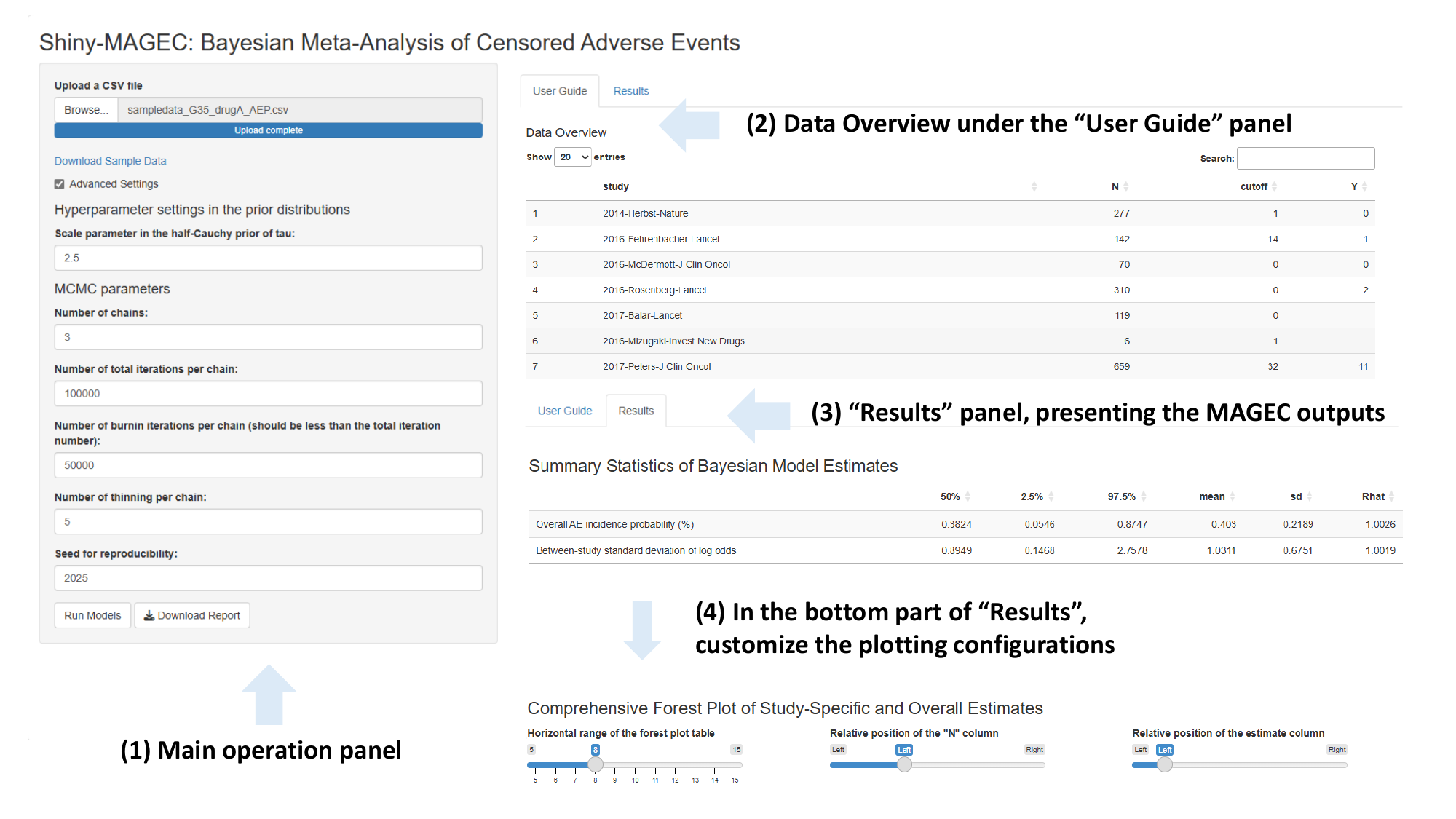}
    \caption{An overview of the operation and result panels in Shiny-MAGEC}
    \label{fig: panel_illustrate}
\end{figure}

In this section, we provide a walk-through of the application, outlining its main features and functions. Users may refer to Section \ref{sec:example} for an illustrative example using a sample data excerpted from the real data application initially investigated in \citet{wang2019treatment} and discussed in \citet{xinyue2024meta}. The application consists of an operation panel on the left and a result panel on the right. The right panel is organized into two primary tabs: (1) ``User Guide'', providing basic information about the software and showing an overview of the uploaded dataset, and (2) ``Results'', displaying the analysis outputs. The accepted data format, advanced settings tunable by users, and the result presentations supported by the application are introduced below. The general layout and features are summarized in Figure \ref{fig: panel_illustrate}. 

\subsection{Data Preparation} \label{sec:prep}

To familiarize themselves with the accepted data format, users can download the sample dataset at the operation panel for reference. The sample data, a subset of the AE meta-analysis dataset from \citet{wang2019treatment}, is displayed in Table \ref{tb: real data}. The input data should be provided as a CSV file with mandatory column names: ``study'', ``N'', ``Y'', and ``cutoff''. The ``study'' column is a unique identifier for each study. We recommend using the abbreviated names of different studies (i.e., character strings) to label the studies since these labels will be used in presenting and plotting the study-specific results as shown in Section \ref{sec:operation}. The ``N'' column is the sample size assigned to the treatment in each study, and the ``Y'' column (when not missing) represents the observed AE count given the prespecified severity grade and category of interest. ``Y'' should be coded as \texttt{NA} or left blank (preferred) when an AE count is unreported. The ``cutoff'' column gives the study-specific left-censoring threshold, which can be extracted from a footnote or methods description in the original publication. Each cutoff is the largest integer not reported; for example, in a study with $N=459$ and a footnote that ``AE counts $\ge2\%$ of the treated size are disclosed,'' the study-specific cutoff would be $9$, since any count larger than $459\times 2\%=9.18$ would be reported.

After uploading the meta-analysis data following the required format, the application will automatically display the data content in the ``Data Overview'' section in the ``User Guide'' tab for a routine check. This immediate feedback allows users to verify that their data have been read correctly (e.g., all studies are listed and the columns are interpreted as intended) before proceeding.

\begin{table*}[!ht]
        \caption[]{An example AE data subset from \citet{wang2019treatment} including the grade 3 to 5 pneumonitis counts for patients treated with Atezolizumab. ``-'' indicates unreported (left-censored) in the original publication. The left-censored cutoffs are calculated specific to different studies. Given a cutoff of 0 (e.g., in \textit{2018-Colevas-Ann Oncol}), the actual pneumonitis count, though unreported, was exactly 0.\\ 
        }       \label{tb: real data}
        \begin{tabular*}{\textwidth}{@{\extracolsep{\fill}}llcccccc@{}}
        \hline
        & Study Source & No. of Treated Patients ($N$) & Cutoff & Pneumonitis Count ($Y$) & \\
        \hline
        & 2014-Herbst-Nature & 277  &  1 &  0 & \\
        & 2016-Fehrenbacher-Lancet & 142 &  14 &  1 & \\ 
        & 2016-McDermott-J Clin Oncol & 70 &  0 &  0 &\\
        & 2016-Rosenberg-Lancet & 310 &  0 &  2 &\\
        & 2017-Balar-Lancet & 119 &  0 &  - &\\
        & 2016-Mizugaki-Invest New Drugs & 6 &  1 & - &\\ 
        & 2017-Peters-J Clin Oncol & 659 & 32 & 11 &\\ 
        & 2017-Rittmeyer-Lancet & 609 & 60 &  4  &\\
        & 2018-Colevas-Ann Oncol & 32 & 0 &  -  &\\
        & 2018-Emens-JAMA Oncol & 116 & 3 &  1  &\\
        & 2018-Horn-Eur J Cancer & 89 & 4 &  0  &\\
        & 2018-Lukas-J Neurooncol & 16 & 0 &  -  &\\
        & 2018-McDermott-Nat Med & 103 & 20 &  0  &\\
        & 2018-Petrylak-JAMA Onc & 95 & 0 &  -  &\\
        & 2018-Powles-Lancet & 459 & 9 &  -  &\\
        \hline
        \end{tabular*}
\end{table*}

\subsection{Operation and Results of Shiny-MAGEC} \label{sec:operation}
Once the data are loaded, the analysis can be initiated by clicking the ``Run Models'' button. Two Bayesian models will be sequentially executed: one that incorporates the censored data and uses the MAGEC approach and another that solely utilizes the fully observed data. The latter serves as an optional supplementary reference highlighting the bias that might be caused by the inappropriate complete-case analysis procedure. The Bayesian model estimations are based on MCMC simulations, and the sampling algorithm is the Metropolis-Hasting algorithm implemented using the JAGS software (Version 4.3.1). A few advanced parameters regarding the prior specifications in the models and the MCMC simulations can be manipulated by checking the box of ``Advanced Settings''. By default, the scale parameter $A$ in the prior distribution of $\tau$ described in Section \ref{sec:model} is set to $2.5$. For MCMC, three chains are run in parallel, each with a total of 100,000 iterations, that include a burn-in of 50,000 and a thinning interval of 5 (i.e., keeping 1 out of every 5 samples). These defaults enhance robust estimation and an acceptable convergence performance in general applications. 

When the model fitting is finished, the application generates tabular and figure outputs for the MAGEC model shown in the ``Results'' tab. Within the ``Results'' tab, first, summary statistics about the posterior estimates of the overall AE incidence probability and the between-study variation are shown in a table. This includes the posterior median, the standard deviation (SD), boundaries of the $95\%$ credible interval (CrI), and the mean and standard error of the posterior distributions. It is followed by a short paragraph briefly summarizing the key meta-analytic results. For the purpose of comparison, the biased overall AE incidence probability from the complete-case analysis is also provided. The summary table also includes the Gelman-Rubin potential scale reduction factors \texttt{Rhat}s \citep{gelman2013bayesian} for diagnosing the convergence of the MCMC chains. If the \texttt{Rhat} of any key parameters is larger than $1.01$, a warning will be output, suggesting the user increase the lengths of the MCMC chains in the ``Advanced settings'' to improve the convergence and mixing performance. 

In the bottom part of the page, the meta-analytic estimates as well as the study-specific estimates will be comprehensively visualized in a forest plot. Users may navigate to the advanced operation panels on the tab to customize the figure. Finally, an analysis report in Microsoft Word format containing the summary tables and the forest plots of both the MAGEC analysis and the complete-case analysis can be downloaded by clicking the ``Download Report'' button on the operation panel. To facilitate scientific writing and reporting, users can also find a template for drafting the statistical analysis section in the downloaded analysis report, which details the statistical methods for the MAGEC modeling (e.g., the specification of MCMC simulations) and includes relevant references.

\section{An Illustrative Example} \label{sec:example}

We present an illustrative example to demonstrate the use of Shiny-MAGEC based on the built-in sample dataset. \citet{wang2019treatment} conducted a meta-analysis of 125 clinical studies evaluating the incidence probabilities of pneumonitis (a type of AE involving inflammation of lung tissue) for several types of PD-1 and PD-L1 inhibitors. As shown in Table \ref{tb: real data} (Section \ref{sec:prep}), our sample data is a small subset of the full data focused on the incidence of Grade 3 to 5 pneumonitis for patients that received Atezolizumab.

Once the data is uploaded, a data overview can be found in the ``User Guide'' tab. By clicking the ``Advanced Settings'' checkbox on the left, a panel allowing customization of prior hyperparameters and MCMC options will show up. In this illustrative example, we use the default settings. After clicking the ``Run Model'' button, a progress bar will approximately indicate the models' execution statuses. In this illustrative example, the running time on an AMD R9-6900HS CPU is approximately 10 seconds. Once the posterior sampling is completed, we can navigate to the ``Results'' page to review the model outputs.

\begin{figure}[H]
    \centering
    \includegraphics[width=\linewidth]{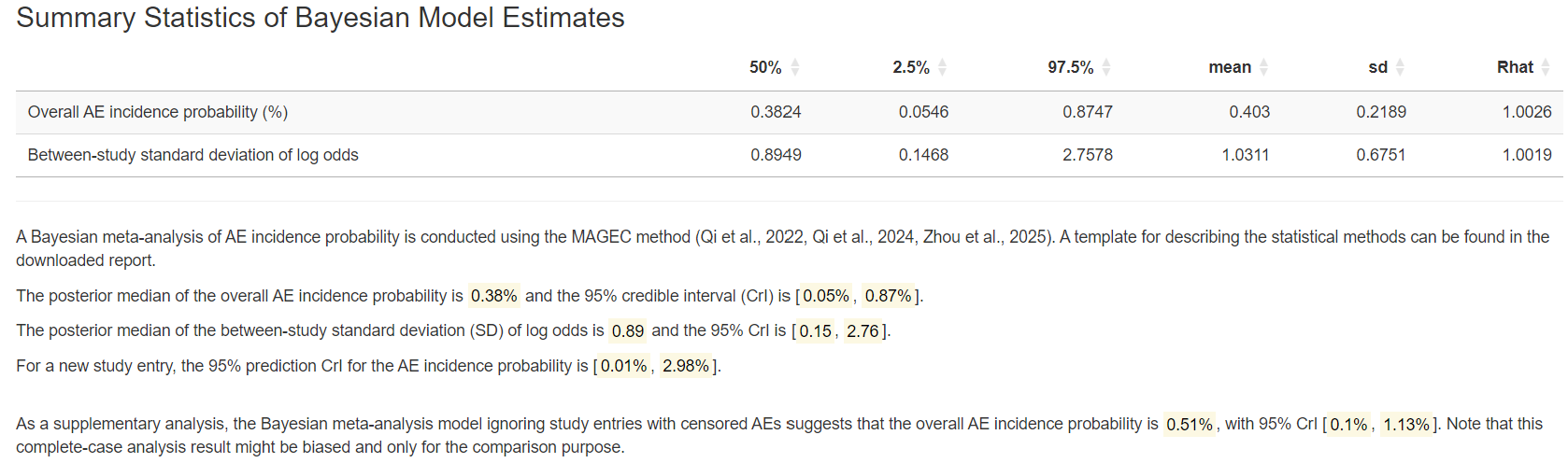}
    \caption{Shiny app outputs of the summary statistics table and result descriptions based on the illustrative example. 
    }
    \label{fig: sumtable}
\end{figure}

The main results generated based on the MAGEC methodology are displayed in the ``Results'' tab, while the parallel results derived from the complete-case analysis are summarized briefly in a separate paragraph only for supplementary comparison purposes. Figure \ref{fig: sumtable} illustrates the summary statistics of the MAGEC model estimates, along with an automatically generated paragraph summarizing the key meta-analytic results. Users can customize this generated draft further to match their specific study contexts. 

\begin{figure}[H]
    \centering
    \includegraphics[width=\linewidth]{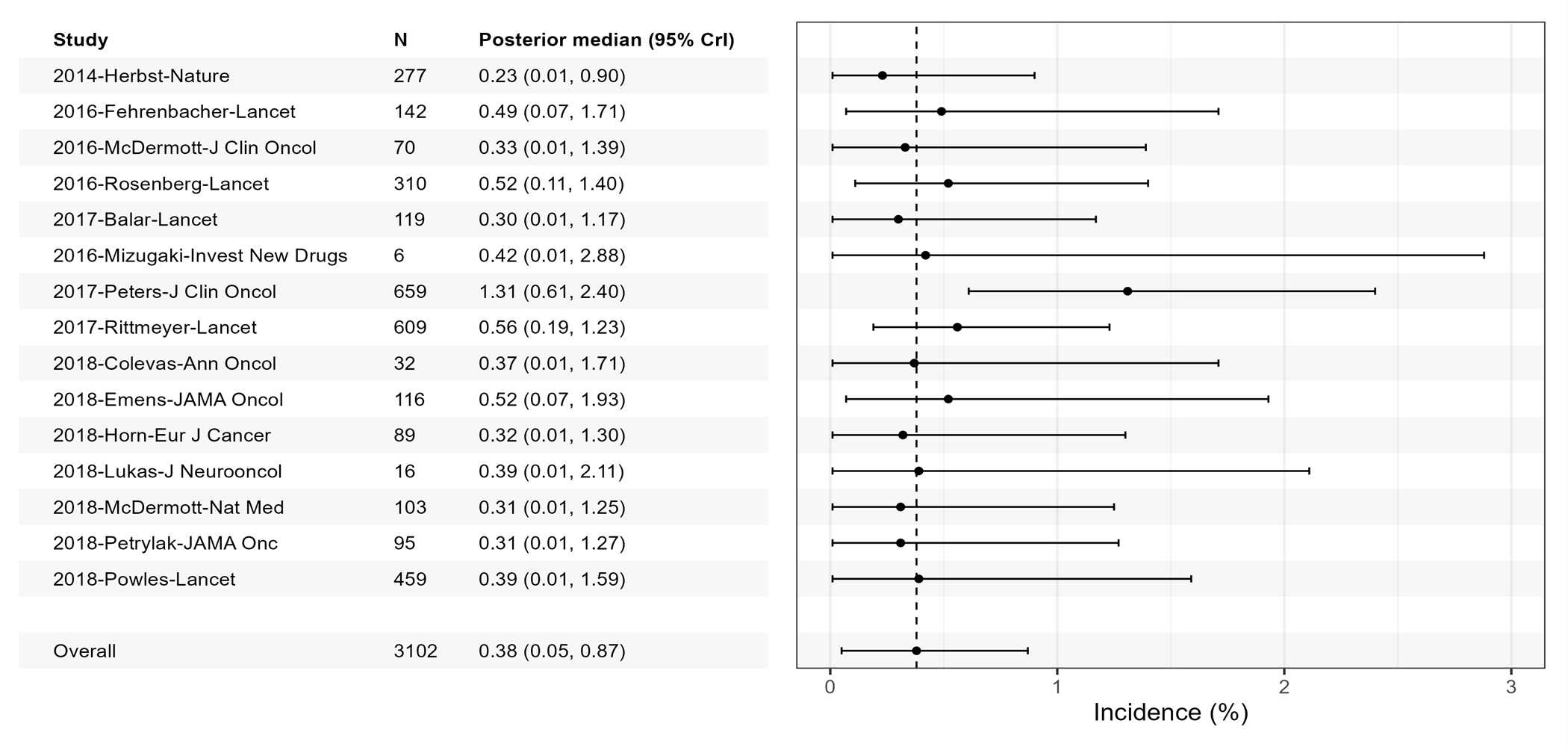}
    \caption{Comprehensive forest plot of the incidence probabilities (in percentage) of Grade 3 to 5 pneumonitis based on the MAGEC meta-analysis model.}
    \label{fig: forestplot}
\end{figure}

Additionally, a forest plot will be generated to illustrate both the study-specific and overall AE incidence probability estimates (in percentages), as presented in Figure \ref{fig: forestplot}. This visualization will follow established scientific reporting structures exemplified in prior literature, including works by \citet{wang2019treatment} and \citet{xinyue2024meta}. As demonstrated in prior research, this presentation approach provides a clear and structured summary of results, enhancing interpretability and comparability of the results.

It is straightforward that by incorporating richer information provided by the censored studies, the MAGEC model provides different point and interval estimates compared to the complete-case analysis that solely uses the non-censored studies.  By applying the MAGEC modeling approach, the AE incidence is estimated at 0.38\% (95\% Crl [0.05\%, 0.87\%]) in this example, whereas the complete case analysis yields a higher estimate of 0.51\% (95\% Crl [0.10\%, 1.13\%]). The discrepancy accounts for a 34\% over-estimation bias inherent in the complete case approach, reinforcing findings from previous studies that have shown complete case methods tend to inflate incidence estimates due to missing data \citep{xinyue2024meta}.  

\section{Discussion} \label{sec:discussion}
Meta-analysis has become an indispensable tool for characterizing the safety profiles of medications, but it comes with unique challenges compared to efficacy analysis. Incomplete reporting of AEs, together with possible heterogeneity across studies, can compromise the validity of a drug safety meta-analysis if not properly handled. On the methodological front, researchers are developing more sophisticated models to tackle these challenges. The Bayesian censored-data approach by \citet{xinyue2024meta} is one such innovation that addresses partially observed safety data. In this paper, we extend these efforts by providing a user-friendly tool that brings advanced methodology to a broader audience of clinicians and researchers. 

The Shiny-MAGEC app offers an interactive platform to facilitate the application of the Bayesian meta-analysis method described for censored AE data. By incorporating reporting cutoff information into the analysis, it mitigates the estimation bias of AE incidence rates, accompanied by appropriate uncertainty quantification. This new tool facilitates rigorous systematic review and meta-analysis of safety data by making state-of-the-art bias correction accessible to users without requiring advanced statistical programming skills.

Beyond this specific tool, our work aligns with a broader movement to improve the quality of AE meta-analyses. Regulators (e.g., FDA, EMA) and guideline groups (e.g., Cochrane, CONSORT, PRISMA) have increasingly emphasized the need for high standards in safety evidence synthesis to support robust conclusions about drug risks \citep{us2015adverse,ioannidis2004better,CIOMS_MetaAnalysis2016,higgins2019cochrane,page2021prisma}. A balanced, evidence-based understanding of drug harms is critical to weighing benefits against risks in clinical decision-making. Particularly, regulatory agencies rely on comprehensive safety evidence when making approval and labeling decisions, as flawed or incomplete safety meta-analyses can have serious consequences \citep{us2018guidance,EMA_Guidelines}. Therefore, safety outcomes should be synthesized with the same rigor and transparency as efficacy data, ensuring scientific completeness and reliability in assessments of drug risk. An understated risk in a meta-analysis might delay regulatory actions on a harmful drug, whereas an overestimated risk could unnecessarily alarm practitioners and patients. In practice, both underestimation and overestimation of risks must be avoided. 

While the challenges in meta-analysis of drug safety are non-trivial, they are surmountable with diligent methodology innovation and improved data practices \citep{fda2020}. By implementing best practices for AE reporting, leveraging advanced statistical techniques like the Bayesian models employed here, and exploring new tools like natural language processing for data gathering \citep{young2019systematic}, the field can move toward more reliable and informative safety meta-analyses. These improvements will ultimately benefit patients and healthcare providers, as treatment decisions can be made with a clearer understanding of the balance between benefits and harms. 

The future of drug safety meta-analysis is moving toward greater scientific rigor, transparency, and integration of diverse data sources – all aimed at safeguarding public health through better evidence on drug risks. Continued methodological research and consensus-building are needed to refine how we pool and interpret adverse event data. As tools like our R Shiny app become integrated into researchers’ workflows, we anticipate more accurate and trustworthy assessments of drug safety that will support better-informed clinical and regulatory decisions.

\end{document}